\documentclass[10pt,a4paper]{article}
\usepackage{jheppub}

\usepackage{multirow}
\usepackage{mathrsfs}
\usepackage{amssymb}
\usepackage{verbatim}
\usepackage{rotating} 

\usepackage{bbm}
\usepackage{array}
\usepackage{float}
\usepackage{color}
\usepackage{graphicx}

\newcommand{\Dbkg}{D_{\text{bkg}}}
\newcommand{\Dtest}{D_{\text{test}}}

\newcommand{\dms}{\Delta m^2_{21}}
\newcommand{\dml}{\Delta m^2_{31}}

\newcommand{\bea}{\begin{eqnarray}}
\newcommand{\eea}{\end{eqnarray}}
\newcommand{\be}{\begin{equation}}
\newcommand{\ee}{\end{equation}}

\newcommand{\mbTh}{\mathbf{\Theta}}

\newcommand{\lhood}{\mathcal{L}}
\newcommand{\ev}{\mathcal{Z}}
\newcommand{\NME}{\mathcal{M}}
\newcommand{\mcA}{\mathcal{A}}
\newcommand{\mcB}{\mathcal{B}}
\newcommand{\mcR}{\mathcal{R}}

\newcommand{\keV}{\,\text{keV}}
\newcommand{\eV}{\,\text{eV}}
\newcommand{\yr}{\,\text{yr}}

\newcommand{\Ge}{\text{Ge}}
\newcommand{\Xe}{\text{Xe}}
\newcommand{\DXe}{D_\text{Xe}}
\newcommand{\DGe}{D_\text{Ge}}
\newcommand{\Xew}{$^{136}$Xe}
\newcommand{\Gew}{$^{76}$Ge}

\def\T2K{{\sc T2K }}

\def\MN{{\sc MultiNest}}

\def\U{{\rm U}}
\def\LOG{{\rm LOG}}
\def\NOR{{\rm N}}
\def\LOGN{{\rm LOGN}}

\newcommand{\ie}{\emph{i.e.}}
\newcommand{\eg}{\emph{e.g.}}

\newcommand{\df}{{\rm d}}
\newcommand{\qu}[1]{``#1''}
\newcommand{\eref}[1]{Eq.~(\ref{#1})}
\def\lbrac{\left\lbrace}
\def\rbrac{\right\rbrace}

\newcommand{\refcite}[1]{Ref.~\cite{#1}}

\newcommand{\figref}[1]{Fig.~\ref{#1}}
\newcommand{\tabref}[1]{Tab.~\ref{#1}}
\newcommand{\secref}[1]{Sec.~\ref{#1}}

\title{Combining and comparing neutrinoless double beta decay experiments using different nuclei}
\author{Johannes Bergstr\"om}
\affiliation{Department of Theoretical Physics, School of Engineering Sciences\\
KTH Royal Institute of Technology -- AlbaNova University Center\\
Roslagstullsbacken 21, 106 91 Stockholm, Sweden}
\emailAdd{johbergs@kth.se}

\abstract{
We perform a global fit of the most relevant neutrinoless double beta decay experiments within the standard model with massive Majorana neutrinos. Using Bayesian inference makes it possible to take into account the theoretical uncertainties on the nuclear matrix elements in a fully consistent way. First, we analyze the data used to claim the observation of neutrinoless double beta decay in \Gew, and find strong evidence (according to Jeffrey's scale) for a peak in the spectrum and moderate evidence for that the peak is actually close to the energy expected for the neutrinoless decay. We also find a significantly larger statistical error than the original analysis, which we include in the comparison with other data. Then, we statistically test the consistency between this claim with that of recent measurements using \Xew. We find that the two data sets are about 40 to 80 times more probable under the assumption that they are inconsistent, depending on the nuclear matrix element uncertainties and the prior on the smallest neutrino mass. Hence, there is moderate to strong evidence of incompatibility, and for equal prior probabilities the posterior probability of compatibility is between $1.3 \%$ and $2.5 \%$. 
If one, despite such evidence for incompatibility, combines the two data sets, we find that the total evidence of neutrinoless double beta decay is negligible. If one ignores the claim, there is weak evidence against the existence of the decay. We also perform approximate frequentist tests of compatibility for fixed ratios of the nuclear matrix elements, as well as of the no signal hypothesis. Generalization to other sets of experiments as well as other mechanisms mediating the decay is possible.
}
\begin{document}
\maketitle

\section{\label{sec:intro}Introduction}
The well-established phenomenon of neutrino oscillations requires that at least two of the three neutrinos of the standards model are massive, and massive neutrinos could cause observable effects in other experiments. Electron spectra from beta decaying nuclei could be affected, but searches for such modifications have so far come up negative \cite{Kraus:2004zw,Aseev:2011dq}. Furthermore, cosmological observations can also provide information, although the constraints depend on which cosmological model is assumed \cite{Komatsu:2010fb,GonzalezGarcia:2010un}.

Since the neutrinos do not have any charges under known unbroken gauge symmetries, it is possible that the neutrinos are Majorana particles, \ie, their own antiparticles. The most efficient way of determining this is to search for a certain kind of nuclear decay in which a nucleus undergoes a double beta decay without emitting any neutrinos, so-called \emph{neutrinoless double beta decay}. There are numerous particles in renormalizable models beyond the standard model which could mediate the decay \cite{Rodejohann:2011mu}, and one can also perform studies using effective field theories \cite{Pas:2000vn,Bergstrom:2011dt}. However, the most commonly studied fields mediating the decay are the standard model neutrinos with Majorana masses. In this case, neutrinoless double beta decay is sensitive to the effective mass given by
\be
m_{ee} \equiv |(\mathcal{M}_\nu)_{ee}| = \left| \sum_i U_{ei}^2 m_i \right| = |m_1 c_{12}^2 c_{13}^2 + m_2 s_{12}^2 c_{13}^2 e^{2i\alpha} + m_3 s_{13}^2 e^{2i\beta}|,
\ee
where $\mathcal{M}_\nu$ is the Majorana mass matrix of the neutrinos and $(\mathcal{M}_\nu)_{ee}$ is its $ee$'th element. The $m_i$'s are the masses of the mass eigenstate neutrinos, $c_{ij}$ and $s_{ij}$ are the cosines and sines of the mixing angles $\theta_{ij}$, and $\alpha$ and $\beta$ are the so-called Majorana phases.
The inverse half-life of a given nucleus $N$ due to neutrinoless double beta decay is given by
\be \label{eq:decayrate}
T_{N}^{-1} = G_{N} |\mathcal{M}_{N}|^2 m_{ee}^2,
\ee
with $G_{N}$ a known numerical \emph{phase space factor}. $\NME_N$ is the \emph{nuclear matrix element} (NME), which encodes the nuclear processes at work. Its calculation is a very difficult problem, and requires certain approximations and assumptions to be made.
There are different models and approaches used to calculate the NMEs, which have often given quite different results, although much progress has been made in recent years. As a consequence, the required values of the NMEs have rather large \qu{theoretical} uncertainties, which needs to be taken into account when analyzing data within models beyond the standard model predicting the decay. For further information, see, \eg, Refs.~\cite{Rodejohann:2011mu,GomezCadenas:2011it,Bilenky:2012qi}.

There has been no clear detection of the neutrinoless decay generally accepted by the community, although a small subset of the Heidelberg-Moscow experiment, using \Gew~as the decaying nucleus, did make such a claim \cite{KlapdorKleingrothaus:2002md}. This analysis was subsequently updated in Refs.~\cite{KlapdorKleingrothaus:2004wj,KKK} (see also \refcite{Kirpichnikov:2010nu}). For many years, there was no other experiment with enough sensitivity to test these claims. Recently, however, the first data from a new generation of experiments has been released \cite{Auger:2012ar,Gando:2012zm}, which have sensitivities in the same region of $m_{ee}$-values preferred by the \Gew~claim, and which show no evidence of the decay. However, these experiments use \Xew~as the decaying nucleus, which has its own NME, also with a large uncertainty. In order to combine and compare these results, one first needs to chose a specific underlying mechanism predicting the decay, and take into account the uncertainties on both the NMEs. 

The aim of this work is to perform such an analysis within the standard model with massive Majorana neutrinos, and we choose to concentrate on a  Bayesian analysis since this, in addition to all its usual advantages (see, \eg, \refcite{Hobson:2010book}), allows us to take into account the large NME uncertainties in a fully consistent and statistically coherent way. Often, one is interested in obtaining constraints using a combination of multiple sets of data. However, one should first be convinced that the different data sets are actually consistent within the model under study. Testing the consistency is extra relevant in the case of neutrinoless double beta decay since the recent data seems to be rather inconsistent with the claim of \refcite{KKK}. Our analysis will take into account the statistical uncertainties in all the data sets as well as the theoretical uncertainties of the NMEs, neither of which was fully considered in the comparisons of Refs.~\cite{Auger:2012ar,Gando:2012zm}.

This work is organized as follows. In Sec.~2, we review the principles of Bayesian
inference, with focus on Bayesian model selection and the compatibility test we will employ. Sec.~3 is an analysis of the data presented in \refcite{KKK}, which is necessary in order to make the comparison with the other data sets possible. Sec.~4 is a description of the other data and likelihoods used in the analysis, while Sec.~5 describes the underlying model, its parameters, the treatment of the NME uncertainties, and the priors used. In Sec.~6, the results of the analysis are presented, and the conclusions can be found in Sec.~7.

\section{Bayesian inference }\label{sec:BayesInf}
In the Bayesian interpretation, probability is associated with degree of belief.
This is in contrast to the frequentist interpretation, in which probability is defined as the limit of the relative frequency of an event in a large number of repeated trials.

If one accepts the Bayesian interpretation, a very powerful arsenal of inference tools become available. In essence, Bayesian inference is a framework for updating prior belief or knowledge based on new information or data.  
Generally, the probability $\Pr(A | B)$ represents the degree of belief regarding the truth of $A$, given $B$. The order of the conditioning can be reversed using \emph{Bayes' theorem},
\begin{equation} \Pr(A|B) =
\frac{\Pr(B|A)\Pr(A)}{\Pr(B)}.
\end{equation}

Perhaps the central purpose of science is to infer which model or hypothesis best, and usually most economically, describes a certain set of collected data. If the collected data is denoted by $\mathbf{D}$ and the set of plausible hypotheses $\lbrac H_i \rbrac_{i=1}^r$, the most straightforward Bayesian solution is to simply use Bayes' theorem to calculate the \emph{posterior probability} of each of the hypotheses,\footnote{All probabilities are also implicitly assumed to be conditioned on all the relevant background information $I$, \ie, $\Pr(X)$ is written instead of $\Pr(X|I)$.} 
\begin{equation}\label{eq:Bayes_model} \Pr(H_i|\mathbf{D}) = \frac{\Pr(\mathbf{D}|H_i)\Pr(H_i)}{\Pr(\mathbf{D})} = \frac{\Pr(\mathbf{D}|H_i)\Pr(H_i)}{\sum_{j=1}^r \Pr(\mathbf{D}|H_j)\Pr(H_j)}.
\end{equation} 
Here, $\ev_i \equiv \Pr(\mathbf{D}|H_i)$ is the probability of the data, assuming the model $H_i$ to be true, and is often called the \emph{evidence} of the model $H_i$. Equation \ref{eq:Bayes_model} can then also be written as
\begin{equation}\Pr(H_i|\mathbf{D}) = \frac{1}{1+\sum_{i \neq j}\frac{\mathcal{Z}_i}{\mathcal{Z}_j} \frac{\Pr(H_i)}{\Pr(H_j)}}.\label{eq:Bayes_model_2}\end{equation} 
If one is not comfortable with assigning absolute probabilities to the different hypothesis, one can instead consider only the \emph{posterior odds}, which is the ratio of the posterior probabilities,
\begin{equation}\label{eq:post_ratio} \frac{ \Pr(H_i|\mathbf{D})}{\Pr(H_j|\mathbf{D})} =
\frac{\Pr(\mathbf{D}|H_i)}{\Pr(\mathbf{D}|H_j)} \frac{\Pr(H_i)}{\Pr(H_j)} = \frac{\mathcal{Z}_i}{\mathcal{Z}_j} \frac{\Pr(H_i)}{\Pr(H_j)},
\end{equation}
which implies that the posterior odds equals the prior odds (often chosen as unity) times the \emph{Bayes factor}, the ratio of the evidences. This method of comparing models is usually called \emph{model selection}, although \emph{model comparison} or \emph{model inference} might be more accurate descriptions in the case that no single model is actually selected.

If the model $H$ is \emph{simple}, \ie, has no free parameters, then the evidence is simply the probability (density) of the data $\mathbf{D}$ when $H$ is assumed to be true. 
If instead the model contains free parameters $\mbTh$, straightforward application of the laws of probability implies that the evidence is given by
\bea
\mathcal{Z} =\Pr(\mathbf{D}|H) &=&  \int \Pr(\mathbf{D},\mbTh|H)\df^N\mathbf{\Theta} = \int \Pr(\mathbf{D}|\mbTh, H) \Pr(\mathbf{\Theta}|H)\df^N\mathbf{\Theta} \notag \\
&=& \int{\mathcal{L}(\mathbf{\Theta})\pi(\mathbf{\Theta})}\df^N\mathbf{\Theta}.
\label{eq:Z}
\eea
Here, the \emph{likelihood function} $\mathcal{L}(\mathbf{\Theta}) \equiv \Pr(\mathbf{D}|\mathbf{\Theta}, H) $ is the probability (density) of the data $\mathbf{D}$, assuming parameter values $\mbTh$ and $\pi(\mathbf{\Theta}) \equiv \Pr(\mathbf{\Theta}|H)  $ is the prior probability (density), which should reflects one's degree of belief of the parameters, given the model and the background information but not the data. 

One observes that the evidence is the average of the likelihood over the prior, and hence this method automatically implements a form of \emph{Occam's razor}, since in general a more predictive model will have a larger evidence than a less predictive one, unless the latter can fit the data substantially better. Bayes factors or posterior odds are usually interpreted using \emph{Jeffrey's scale} in \tabref{tab:Jeffreys}\footnote{We denote the natural logarithm as \qu{$\log$} and the base-10 logarithm as \qu{lg}.}, as used in, for example, Refs.~\cite{Trotta:2008qt,Trotta:2005ar,Hobson:2010book,Feroz:2008wr,AbdusSalam:2009tr}. Note that probability itself implies a somewhat unique and meaningful scale of the evidence, and that \tabref{tab:Jeffreys} simply gives rough descriptive statements of posterior odds and probabilities.

\begin{table}
\begin{center}
\begin{tabular}{|c|c|c|c|}
\hline
$\log(\text{odds})$ & odds & $\Pr(H_1 | \mathbf{D})$ & Interpretation \\ 
\hline
$<1.0$ & $\lesssim 3:1$ & $\lesssim 0.75$ & Inconclusive \\
$1.0$ & $\simeq 3:1$ &  $\simeq 0.75$ & Weak evidence \\
$2.5$ & $\simeq 12:1$ & $\simeq 0.92$ & Moderate evidence \\
$5.0$ & $\simeq 150:1$ & $ \simeq 0.993$ & Strong evidence \\ \hline
\end{tabular}
\end{center}
\caption{Jeffrey's scale often used for the interpretation of Bayes factors, odds, and model probabilities. The posterior model probabilities for the preferred model are calculated by assuming only two competing hypotheses and equal prior probabilities.}
\label{tab:Jeffreys}
\end{table}

Note that it is often the case that the evidence is quite dependent on the prior used \cite{Kass:1995,Trotta:2005ar}, although the Bayes factor will generally favour the correct model once \qu{enough} data has been obtained. Furthermore, when comparing nested models, taking the Bayesian view also means that the significance, or the \qu{number of $\sigma$'s}, of a result is in general not a good indicator of the importance or the evidence of a new effect, a result that is known as \qu{Lindley's paradox}. For further details, see, \eg, Appendix A of \refcite{Trotta:2005ar}.

Once one or a set of models with large posterior probabilities has been found, the complete inference of the parameters of those models are given by the posterior distribution through Bayes' theorem,
\begin{equation} \Pr( \mathbf{ \Theta} | \mathbf{D},H) = \frac{\Pr(\mathbf{D}
|\mathbf{\Theta},H)\Pr(\mathbf{\Theta}|H)}
{\Pr(\mathbf{D}|H)}  = \frac{\lhood(\mbTh)\pi(\mbTh)}{\ev}.
\end{equation}
Since the evidence does not depend on the values of the parameters $\mbTh$, it is usually ignored in parameter estimation problems and the parameter inference is obtained using the unnormalized posterior. However, note that the evaluation of the posterior distribution of the parameters is only meaningful if the model does not have a very small posterior probability, since otherwise the model as a whole is strongly disfavored. In practise, this means that one should \emph{first} calculate the evidences and posterior odds and only then, for the models with not to small evidences, calculate the posterior distribution.

In fact, if more than one model has a significant probability, it is better to consider the distributions of parameters not assuming the model with maximum probability to be correct, but instead take into account the uncertainty regarding which model is the correct one, giving the model-averaged distribution \cite{Kass:1995}
\be  \Pr(\mathbf{\Theta} |X) = \sum_{i=1}^r \Pr(\mathbf{\Theta} |H_i, X) \Pr(H_i |X), \ee
which is the probability distribution given by the average of the individual distributions over the space of models, with weights equal to the  model probabilities. 

The main result of Bayesian parameter inference is the posterior and its marginalized versions (usually in one or two dimensions).
However, it is also common to give point estimates such as the posterior mean or median, as well as \emph{credible intervals (regions)}, which are defined as intervals (regions) containing a certain amount of posterior probability. Note that these regions are not unique without further restrictions, just as for classical confidence intervals, and that in general they do not contain all the information that the posterior contains.

Although the reasoning and techniques used when performing model selection are often different than when estimating parameters, one can equally well consider model selection as a parameter inference problem with an additional discrete parameter denoting the model index. Hence, there is no real \qu{fundamental} difference between model selection and parameter estimation. We use \MN~\cite{Feroz:2007kg,Feroz:2008xx} for the evaluation of all evidences and posterior distributions in this work.

\subsection{A Bayesian consistency test\label{sec:MScomptest}}
One can also use Bayesian model selection to test if a set of data is consistent or not within a given model $H$ \cite{Marshall:2004zd,Feroz:2008wr,Feroz:2009dv}.  
First, one partitions the data as $\mathbf{D} = (\Dtest,\Dbkg)$, where we want to test the internal consistency of $\Dtest = (D_1, D_2, \ldots , D_k)$, $k\geq 2$, given $\Dbkg$, a set of possible background data, assumed to be correct and internally consistent.
Let
\begin{itemize}
\item[$C$:] The data $\Dtest$ considered are all consistent within $H$, given the background data $\Dbkg$.
\item[$\bar{C}$:]$\Dtest$ are inconsistent and hence lead to different regions of parameter space being preferred, \ie, $(D_1, D_2, \ldots , D_k)$ need different sets of parameters to describe the data.
\end{itemize} 
We want to calculate the ratio of posterior probabilities of $C$ and $\bar{C}$ (with implicit conditioning on $H$), given by 
\be 
 \frac{\Pr(C|\Dtest, \Dbkg )}{\Pr(\bar{C}|\Dtest, \Dbkg )}  = \frac{\Pr(\Dtest | \Dbkg, C )}{\Pr(\Dtest| \Dbkg , \bar{C})} \frac{\Pr(C|\Dbkg)}{\Pr(\bar{C}|\Dbkg)} = \frac{\Pr(\Dtest | \Dbkg, C)}{\Pr(\Dtest| \Dbkg , \bar{C})}  \frac{\Pr(C)}{\Pr({\bar{C}})} \label{eq:O},
\ee
where we have in the last step have used that $ \Pr(C|\Dbkg) / \Pr(\bar{C}|\Dbkg) = \Pr(C)/\Pr({\bar{C})}$, since the probability that $\Dtest$ is consistent should not change without considering it. From this also follows that $\Pr(\Dbkg|\bar{C}) = \Pr(\Dbkg|C) = \Pr(\Dbkg)$.
The calculable part of \eref{eq:O} is the Bayes factor
\be \mathcal{R} = \frac{\Pr(\Dtest | \Dbkg, C)}{\Pr(\Dtest| \Dbkg , \bar{C})}  =  \frac{\Pr(\Dtest | \Dbkg)}{\prod_{i=1}^k  \Pr(D_i | \Dbkg)}, \label{eq:R_1}\ee
where the last step follows from the defining property of the hypotheses $C$ and $\bar{C}$: the data in $\Dtest$ can be described by the same parameters (of $H$) under $C$, but need different sets under $\bar{C}$.

These are in principle \qu{ordinary} evidence integrals, with the exception that the prior used in the integral is conditioned on $\Dbkg$, \ie, it can be considered the posterior of the data $\Dbkg$. Hence,
\be \Pr(\Dtest | \Dbkg) = \int \Pr(\Dtest | \mbTh) \Pr(\mbTh | \Dbkg) \df^N \mbTh, \ee
and similarly for the other evidences. The conditioning on $\Dbkg$ can be dropped in the likelihood since the probability distribution of the data $\Dtest$ does not depend on $\Dbkg$ if all the free parameters $\mbTh$ are fixed.  If there is no background data, one of course simply uses the \qu{original} priors in the evidence integrals.

However, the integral can be difficult to perform in practise if $\Pr(\mbTh | \Dbkg)$ is not simple.
Then one can use that \eref{eq:R_1} can be written as
\be 
\mathcal{R} = \frac{\Pr(\Dtest,\Dbkg | H)}{\prod_{i=1}^k   \Pr(D_{i},
  \Dbkg | H)}  \Pr(\Dbkg|H )^{k-1}      \label{eq:R_2}.
\ee
These evidences are the evidences using the original priors, but now also including the background data in the likelihoods.

Although the expression for the Bayes factor in Eqs.~(\ref{eq:R_1}) and (\ref{eq:R_2})  (and from that the posterior probability of consistency) is derived from probability theory, it can still be good to test it on problems where it is obvious what the result \qu{must} be. This has been done on both simple and more advanced toy problems in Refs.~\cite{Marshall:2004zd,Feroz:2008wr,Feroz:2009dv} for the special case $k = 2$. As an analytical example, let $D_1$ and $D_2$ result in likelihoods $\lhood_1(\mbTh)$ and $\lhood_2(\mbTh)$, respectively. If one of the likelihoods (say $\lhood_2$) is constant, and hence give no information on the parameter values, one should obtain $\mathcal{R} = 1$ for any background data and $\lhood_1$, since in this case there is only one actual measurement and hence one cannot say anything about the compatibility. This is indeed what \eref{eq:R_2} reduces to. As a second example, consider the case 
when the second data set determines the model parameters exactly, $\lhood_2(\mbTh) = \delta(\mbTh - \mbTh_0)$. In this case the compatibility test should be equivalent to testing if the first data set favors the null hypothesis $\mbTh = \mbTh_0$ or the more complex one with prior $\pi(\mbTh)$, or $\pi(\mbTh | \Dbkg)$ if background data is included. Indeed, one finds that $\mathcal{R} = {\Pr(D_1| \mbTh_0)}/{\Pr(D_1|H, \Dbkg)} = \lhood_{1}(\mbTh_0)/\int \lhood_{1}(\mbTh) \pi(\mbTh| \Dbkg) \df^N \mbTh$.

\section{The data, likelihood, and evidence of the claim using \Gew.}\label{sec:KKKanalysis}
\begin{figure}[t]
\centering
\includegraphics[width=0.8\textwidth]{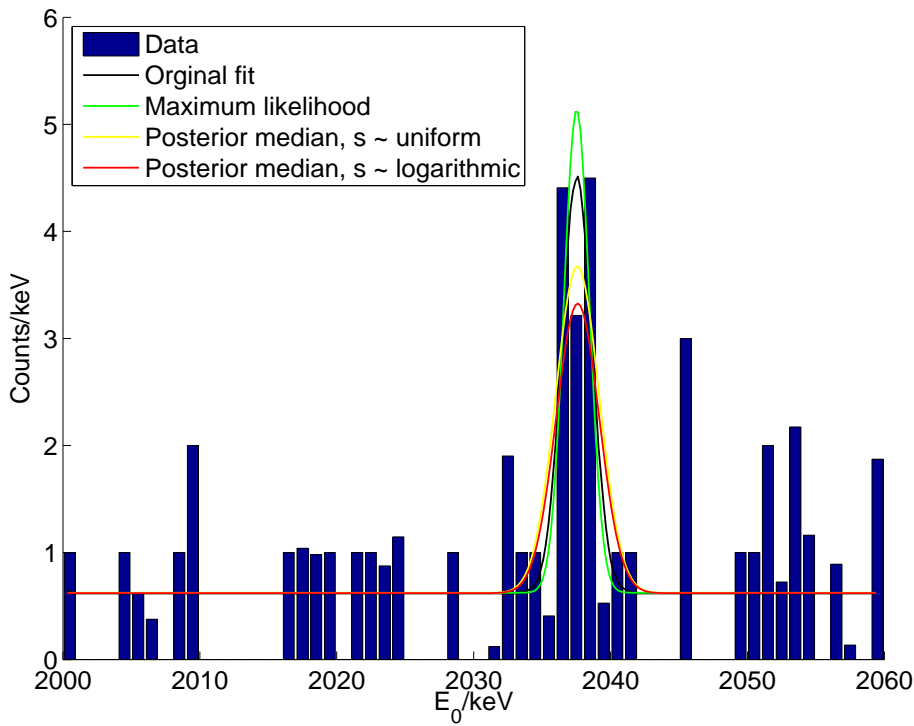}
\caption{The data of Fig.~9 c) of \refcite{KKK}, together with their fit, our maximum likelihood, and posterior median estimates for both the uniform and logarithmic priors on $s$. }\label{fig:spec}
\end{figure}

In order to be able to combine and compare the claim of \refcite{KKK} with other measurements, we will in this section reanalyze the data presented there. There, the authors presented a large number of spectra for different event selection criteria. The choice of which of these to use for further analysis is rather arbitrary; we will use the spectrum in Fig.~9 c) of \refcite{KKK}, which is for the combination of the two event selection methods used, and with the hardest cut on the allowed distance from the edge of the detector. The data, collectively denoted by $\DGe$, is plotted in \figref{fig:spec}.


First, the practise of trying to determine if there is a peak in the spectrum without explicitly considering the null hypothesis of no signal, \ie, implicitly assuming that there is a peak, is not very good. Instead, what one needs to do is to statistically compare the model without a peak with a model which has a peak.
Actually, it seems too restrictive to assume that any possible peak can only come from a neutrinoless double beta decay signal, or something which looks similar. Hence, in addition to the background only model, we consider two signal hypotheses. First, a hypothesis which states that there is a peak \emph{somewhere} in the spectrum, which could for example be a peak from some other radioactive decay with an energy which happens to fall inside the spectrum. Second, a hypothesis stating that there is a signal with the form of that expected from neutrinoless double beta decay, \ie, close to the Q-value $Q=2039.00 \pm 0.05 \keV$ \cite{Douysset:2001cx} of the decay. We assume that any peak has a true width smaller than the energy resolution of the detector. Hence, due to the finite energy resolution of the detector, the expected spectrum in the detector has a peak with the total number of signal events $s$ spread over a distribution (usually taken as a Gaussian) with some width $\sigma$. A related analysis of simulated spectra was performed in \refcite{Caldwell:2006yj}.
  
We thus want to compare the following hypotheses:
\begin{itemize}
\item[$H_0$:] There is only a constant background rate in each bin, which is a priori unknown.
\item[$H_1$:] There is a peak somewhere in the spectrum with unknown position, together with a constant background rate. 
\item[$H_2$:] There is a peak with the properties which is expected of a neutrinoless double beta decay line, and a constant background rate. This means that the peak is centered close to the Q-value of the nucleus.
\end{itemize}
This set of hypotheses could be extended with hypotheses allowing more than one peak in the spectrum, as well as one with a broader peak, but it should be clear from the spectrum in \figref{fig:spec} that these would be disfavored.

A Bayesian analysis of the spectrum in \figref{fig:spec} is in principle straightforward, but is nonetheless an interesting exercise. There is, however, a small complication in the present case. Since the event selection of \refcite{KKK} results in each event being assigned a weight generally different from unity, the resulting spectrum contains non-integer number of \qu{measured events}. The probability distribution in each bin, given an expected number of events, is impossible to estimate, and one notices that there is a much larger probability to have a number of events close to an integer than far away from one. 

For a Bayesian analysis, however, we do not really need the distribution of the data (given values of the parameters), but only the relative probability of the data actually observed as a function of the free parameters. 
We expect this function, the likelihood, to be still reasonably well approximated by\footnote{Note that Poisson likelihoods for non-integer data are sometimes used in high-energy physics \cite{Cowan:2010js}.}
\be \lhood_\Ge(\mbTh) \propto \prod_{i=1}^{n_{\text{bins}}} \lambda_i(\mbTh)^{N_i} e^{-\lambda_i(\mbTh)}, \ee
where $\lambda_i$ and $N_i$ is the predicted and observed number of events in bin $i$, respectively. Under the null hypothesis $\lambda_i = b$, while under the alternatives
\be \lambda_i(\mbTh) = s\int_{E^\text{min}_i}^{E^{\text{max}}_i} \frac1{\sqrt{2\pi}\sigma}e^{-\frac{(E - E_0)^2}{2\sigma^2}} \df E + b,\ee
where bin $i$ is between $E^{\text{min}}_i$ and $E^{\text{max}}_i$, $\mbTh = (s, E_0, \sigma, b)$, and the signal strength, position of the peak, peak width, and background rate are denoted by $s, E_0, \sigma$, and $b$, respectively. 

\begin{figure}[t]
\centering
\includegraphics[width=1.0\textwidth,clip=true]{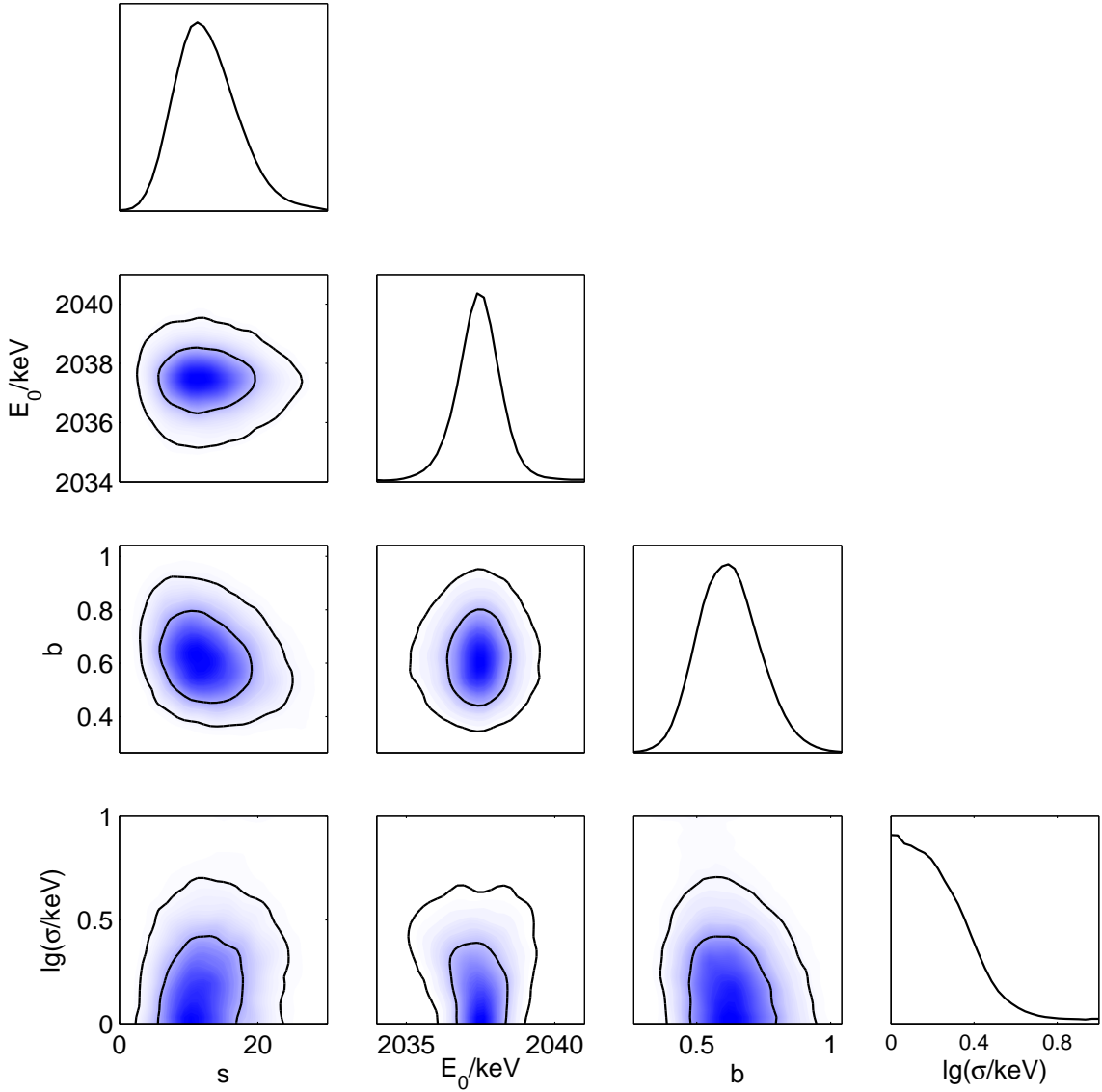}\label{fig:KKKposterior}
\caption{Posterior distributions of the parameters. All the parameters plotted were assigned uniform priors.}
\end{figure}

\begin{table}
\centering
\begin{tabular}{|c|c|c|c|}
\hline
Prior:  & $\log(\ev_1/\ev_0) $  & $\log(\ev_2/\ev_0)$ & $\log(\ev_2/\ev_1) $\\ 
\hline
$s \sim$ $\LOG(0.1,30)$ & $3.37 \pm 0.07$ & $5.81 \pm 0.06$ & $2.44 \pm 0.09$\\ 
$s \sim$  $\LOG(10^{-4},30)$ & $2.55\pm 0.07$  & $4.92 \pm 0.06$ & $2.67 \pm 0.09$\\ 
$s \sim$  $\U(0,30)$ & $4.32 \pm 0.07$  & $6.38 \pm 0.06$ & $2.06 \pm 0.09$\\ 
\hline
\end{tabular}
\caption{\label{tab:logB_KKK}Logarithms of Bayes factors for different priors on the signal strength. The priors on the nuisance parameters are given in \tabref{tab:nuis_priors}. }
\end{table}

\begin{table}[tbh]
\centering
\begin{tabular}{|c|c|}
\hline
Prior: X $\sim$  & Description  \\ 
\hline
$\U(a,b)$ & Uniform between $a$ and $b$ \\ 
 $\LOG(a,b)$ &  $\log(X) \sim \U(\log(a),\log(b))$ \\ 
 $\NOR(\mu,\sigma)$ & Normal with mean  $\mu$ and standard deviation $\sigma$ \\  
 $\LOGN(\mu,\sigma)$ & $\log(X) \sim \NOR(\log(\mu),\log(\sigma))$  \\
\hline
\end{tabular}
\caption{\label{tab:priors} Notation for priors. The uniform prior is zero outside the interval $[a,b]$. For the log-normal prior, the median (but not the mode or mean) of $X$ is $\mu$, while roughly $68.3 \%$ of the prior probability is contained in the interval $[\mu/\sigma, \mu \sigma]$. }
\end{table}

\begin{table}[tbh]
\centering
\begin{tabular}{|c|c|c|c|}
\hline
Model  & $E_0\sim$ & $\sigma \sim $  & $b \sim$ \\ 
\hline
$H_0$ &  $-$ & $-$ & $\U(0,\cdot)$ \\
$H_1$ &  $\U(E^{\text{min}},E^{\text{max}})$ & $\LOGN(1.5 \keV,1.3)$ & $\U(0,\cdot)$ \\
$H_2$ &  $\NOR(2039 \keV, 2 \keV)$ & $\LOGN(1.5 \keV,1.3)$  & $\U(0,\cdot)$ \\
\hline
\end{tabular}
\caption{Priors on nuisance parameters. The notation using a \qu{$\cdot$} as in $\U(0,\cdot)$ means that as long as that limit is chosen large (or small) enough so as to not cut off the posterior, its precise value does not matter.}\label{tab:nuis_priors}
\end{table}

The most important parameter in both $H_1$ and $H_2$ is the signal strength $s$. One might think that a uniform prior on $s$ is the most natural choice. However, one might also be uncertain about the \emph{scale} of the signal, in which case a prior uniform in $\log s$ would be more appropriate. In any case, one needs to specify an upper limit. Neutrinoless double beta decay has been searched for in germanium before \cite{Aalseth:2002rf}, the result of which one can include in the background information and implemented as a (quite conservative) upper limit of $30$ events. The lower limit for the uniform prior is naturally taken to be zero, while for the log priors it should be smaller than about 1 event, but large enough to put a non-negligible amount of prior probability in the region $s \gtrsim 1$. Alternatively, one could assume that the signal would be the result of Majorana neutrino exchange, and use the external constraints from neutrino oscillation and tritium beta decay experiments that exist on $m_{ee}$. Since in this case \eref{eq:decayrate} gives $\log s = C + 2\log \mathcal{M} + 2\log m_{ee}$, $\log s$ would, apart from some smearing due to the uncertainty in the NME, have the same shape of its prior as $m_{ee}$. Reasonable priors for $s$ would then be a logarithmically uniform distribution between roughly $10^{-4} \eV$ and $30 \eV$ for the normal mass ordering and $0.03 \eV$ and $30 \eV$ for the inverted (see \figref{fig:bkg_posterior}). Hence, we use the three different priors of \tabref{tab:logB_KKK},\footnote{The hard lower bound can be avoided by using a prior which decays smoothly, \eg, by taking a uniform distribution below the presently used lower limit, but this will not noticeably change any results.} while the meaning of all the priors used in this work can be found in \tabref{tab:priors}.
The data will also be analyzed within the full model of the standard model with massive neutrinos in \secref{sec:results}.

The main difference between hypotheses $H_1$ and $H_2$ is the prior knowledge put on the position of the peak, $E_0$. Under $H_1$, the most natural choice is to have a uniform prior between the endpoints of the spectrum, $E^{\text{min}} = 2000 \keV$ and $E^{\text{max}} = 2060 \keV$. Under $H_2$ one could simply fix $E_0$ to the Q-value. However, since we by $E_0$ mean the position of the peak in the observed spectrum rather then the true one, one should consider the possibility of a systematic shift of the position of the peak. \refcite{KKK} estimates the systematic uncertainty to about $1.2 \keV$, and in order to allow for this being an underestimation, we take a Gaussian prior of width $2 \keV$, \ie, $E_0 \sim \NOR(2039 \keV, 2 \keV)$. 
Using the information in Refs.~\cite{KlapdorKleingrothaus:2000sn,KlapdorKleingrothaus:2002md,KlapdorKleingrothaus:2004ge,KKK} we estimate the energy resolution to be roughly $1.5 \eV$, but due to lack of detailed information of the experiment, we take a rather conservative uncertainty and $\sigma \sim \LOGN(1.5 \keV,1.3)$. The priors on the nuisance parameters are summarized in \tabref{tab:nuis_priors}. Under all hypotheses, we take a simple uniform prior on the background rate $b$. The resulting posteriors and Bayes factors are insensitive to its upper limit as long as it is large enough, which is denoted by $b \sim \U(0,\cdot)$.

It is always wise to evaluate if the final inference is sensitive to the priors used. The sensitivity to the prior on the signal strength is shown in \tabref{tab:logB_KKK}. For log priors with smaller lower limits, the signal hypotheses become more and more similar to $H_0$, and hence their evidences slowly approach those of $H_0$, but as we can see, this happens very slowly and for reasonable lower limits, this effect is very small and does not change any conclusions. For the nuisance parameters, we have the following. Fixing $\sigma$ at $1.5 \keV$ leaves the log evidence unchanged within the numerical errors of $0.1$, while doubling the uncertainty only lowers the log evidence by $0.3 \pm 0.1$.
In a similar way, one finds that lowering the prior uncertainty of $E_0$ to $1.2 \keV$ (the estimated systematic uncertainty) leaves the evidences invariant within the errors, while increasing it to, say, $3\keV$, only decreases it by about $0.25 \pm 0.1$.

If one assigns equal prior probabilities to the three hypotheses, \eref{eq:Bayes_model_2} together with the Bayes factors of \tabref{tab:logB_KKK} results in the posterior model probabilities 
\bea \Pr(H_0| \DGe) &=& (1.5 - 6.6)\cdot 10^{-3} \\
 \Pr(H_1| \DGe) &=& 0.08-0.11 \\
 \Pr(H_2| \DGe) &=& 0.89-0.92.  \eea
The small variations of the posterior probabilities of $H_1$ and $H_2$ is because $\log(\ev_2/\ev_1)$ is essentially determined by the difference in the prior on $E_0$ and is hence rather independent of the prior on $s$.

The marginalized posteriors of the parameters are shown in \figref{fig:KKKposterior}. They are obtained using the priors of $H_1$, but with the prior on $\sigma$ replaced with a log prior with a lower limit of $1 \keV$, allowing for wider peaks in the spectrum.
The median and $68 \%$ central credible intervals for the signal strength are given by $s = 12.2^{+4.6}_{-3.8}$ and $s = 10.6^{+4.2}_{-3.8}$, for the model $H_2$ with uniform and logarithmic priors on $s$, respectively. However, both quite small and large values of $s$ are allowed, with central $99 \%$ credible intervals given by $[3.7,25.0]$ and $[1.8,23.0]$ for the same two prior choices. The maximum likelihood point and an approximate $68 \%$ confidence interval using the profile likelihood is $s = 10.5^{+4.2}_{-3.5} $. These signal rates and errors should be compared with the reported $s=10.75 \pm 1.58$ of \refcite{KKK}. In \figref{fig:spec} are also plotted the expected distribution of events for different point estimates of the parameters. Our analysis indicate that signal rates a factor of 2 smaller than the best fit, and hence half-lives a factor of 2 larger, are allowed at a reasonable level. We believe that this larger error should be taken into account when comparing with other experiments, which has not been done in previous analyses such as Refs.~\cite{Auger:2012ar,Gando:2012zm}.

Finally, we comment on the significance of a signal using frequentist hypothesis test. The standard hypothesis tests based on the profile likelihood ratio are not applicable because the nuisance parameters are not defined under the null hypothesis (see, \eg, \cite{Protassov:2002sz,Ranucci:2012ed}). In principle one could neglect the uncertainties in $\sigma$ and $E_0$ and fix them, but since the number of events are so small, the expected distribution would most likely be far from the asymptotic one anyway.
In any case, we find
\be Q^2(s = 0) \equiv - 2\log\left(\frac{\sup_{\rho_\Ge} \lhood(s=0,\rho_\Ge )}{\sup_{\mbTh} \lhood(\mbTh)}\right) = 19.8,  \label{eq:Qsquare_Ge} \ee
with $\rho_\Ge= (E_0,\sigma,b)$ the nuisance parameters. Since the distribution of $Q^2$ is not known, this number does not say very much, but for half a $\chi^2$-distribution with 3 and 1 degrees of freedom, the above value of $Q^2$ would correspond to a significance of $3.7$ and $4.4$ standard deviations, respectively. Alternatively, since one knows the expected position of the peak one can also simply merge all bins in in the region $2039 \pm 3 \keV$ (taking into account the energy resolution as well as the systematic uncertainty on $E_0$) into a single counting experiment. One can also imagine that one first fits the background and then fix it to $b\simeq 0.65$. This gives a total of about $n_\Ge^{\text{obs}} = 15$ observed events with an expected background of $4.6$ events in the signal region. Hence, the p-value $\Pr(n_\Ge \geq n_\Ge^{\text{obs}}|H_0) \simeq 9 \cdot 10^{-5}$, corresponding to a $3.7 \sigma$ significance.\footnote{\refcite{Schwingenheuer:2012zs} calculated a $5 \sigma$ significance using a similar method. However, there the observed data was used to select which bins to merge, leading to an overestimated significance.} In addition, the errors on the signal strength is compatible with the naive error estimate of $\sqrt{n_\Ge^{\text{obs}}} \simeq 3.9$. The fact that our Bayesian estimates agrees both with that based on the likelihood and that of naive counting of events, both regarding the point estimates and the errors, strengthens our belief that our analysis is robust.
 
When the data is analyzed within the standard model with massive Majorana neutrinos in \secref{sec:results}, the relation between the inverse half-life of \eref{eq:decayrate} and the signal strength is obtained using the information of \refcite{KKK}, which implies that $s=\epsilon \cdot 2.5 \cdot (T_\Ge / 10^{26} \text{~yr})^{-1}$, with $\epsilon = 1$. According to \refcite{Schwingenheuer:2012zs}\footnote{There is was also pointed out that the error on the signal rate should be of the size we have found.}, this corresponds to assuming a signal detection efficiency of $100 \%$, which is most likely not realistic. Note that a smaller value of epsilon would mean that, in order to reproduce the same signal rate, a larger decay rate, and hence a larger $m_{ee}$ (for fixed NME), would be required.  We will, however, consider the choice $\epsilon = 1$ as part of the considered claim, and only briefly comment on what choosing a smaller $\epsilon$ would imply for the results in \secref{sec:results}.

\section{Other data and likelihoods}
The most relevant other data on neutrinoless double beta decay are the recent measurements using \Xew~as the decaying nucleus \cite{Auger:2012ar,Gando:2012zm}. The EXO collaboration reported one observed event with $4.1$ expected background events within the $\pm 1\sigma$ signal region \cite{Auger:2012ar}. Since the background should be well determined from data outside of this region, we simply take a Poisson likelihood
\be \lhood_{\text{EXO}} \propto  (s_{\text{EXO}}+b_{\text{EXO}})^{N_{\text{EXO}}} e^{-(s_{\text{EXO}} +b_{\text{EXO}} )} \ee
with $N_{\text{EXO}}=1$ and $b_{\text{EXO}}=4.1$. We neglect the uncertainty in $b_{\text{EXO}}$ (reported as $0.3$), although it could easily be incorporated with negligible impact on the end results. The number of signal events is given by
\be s_{\text{EXO}} =  4.7 \cdot \frac{1}{T_\Xe/(10^{25} \text{~yr})}. \ee 
Assuming that the decay is mediated by massive Majorana neutrinos, the half-life is again given by \eref{eq:decayrate}. 

In searches for rare processes in particle physics it is standard practise to not only report frequentist upper limits at some fixed confidence level, but to also report information useful to the rest of the scientific community, making it possible to combine with other experiments and to analyze alternative models. For approximately Gaussian measurements such could be the the maximum likelihood estimate and its error, even if the estimate is for an unphysical value of the parameter. Unfortunately, the KamLAND-Zen Collaboration reports very little information in \refcite{Gando:2012zm} in addition to the observed upper limit. However, with some additional assumptions, one can still obtain an approximate likelihood. Since the expected signal sits on top of a rather large background, the maximum likelihood estimate of the inverse half-life, call it $\hat{\mu}$, should be approximately Gaussian distributed around the true inverse half-life $\mu$. Then, assuming that the collaboration calculated the upper limit using the profile likelihood with the best-fit constrained to be smaller than the tested value (so that only upper limits result, see \refcite{Cowan:2010js}), one can use their stated $90 \%$ sensitivity (median upper limit under the background hypothesis) to estimate the standard deviation of the maximum likelihood estimate as $\sigma_{\hat{\mu}} \simeq 7.8 \cdot 10^{-26} \yr^{-1}$ . In order to estimate the observed maximum likelihood decay rate, we note that \refcite{Gando:2012zm} states that $12 \%$ of hypothetical measurements are expected to yield smaller upper limits under the background hypothesis. Since the upper limits in this case are monotonic functions of the maximum likelihood estimates, one should have $\hat{\mu}^{\text{obs}} \simeq -1.17\sigma_{\hat{\mu}}$. If one summarizes the data with the maximum likelihood estimate, one obtains a Gaussian likelihood with center $\hat{\mu}^{\text{obs}}$ and width $\sigma_{\hat{\mu}}$ as above. In any case, the final results should not depend significantly on reasonable variations of these numbers.
Finally, in order to predict the decay rate in each nucleus in \eref{eq:decayrate}, one needs the phase space factors, which we take from Tab.~1 of \refcite{GomezCadenas:2010gs}, and the nuclear matrix elements, to be discussed in \secref{sec:NMEs}. The EXO and KamLAND-Zen data are collectively denoted by $\DXe$.

Furthermore, we need the likelihoods of the additional constraints on the model resulting from other types of experiments, \ie, the background data $\Dbkg$. These are taken as all neutrino oscillation data analyzed in \refcite{GonzalezGarcia:2012sz} and tritium beta decay data \cite{Kraus:2004zw,Aseev:2011dq}. We do not use cosmological observations nor other double beta decay experiments because of the associated theoretical and model uncertainties.

The neutrino oscillation likelihood is a function of the six oscillation parameters
$ \mathbf{R}_{\rm osc} = (\Delta m_{21}^2,\Delta m_{31}^2, s_{12}^2, s_{23}^2,s_{13}^2, \delta)$.
Since the oscillation parameters (expect $\delta$) are rather well constrained and the correlations between the oscillation in the standard parameterization are rather small, we use an approximation of the likelihood as
\be \lhood_{\rm osc}(\mbTh) \simeq \prod_{i=1}^6 \lhood_{\rm osc}^i(\mathbf{R}_{\rm osc}^i), \ee
where
\be
\lhood_{\rm osc}^{i}(\mathbf{R}_{\rm osc}^i) = \exp\left(-\frac{Q^2_i(\mathbf{R}_{\rm osc}^i)}{2}\right).
\ee 
We do not assume Gaussianity of the individual likelihoods, but instead use the functions $Q_i^2(\mathbf{R}_{\rm osc}^i)$ as plotted in Fig.~2 of \refcite{GonzalezGarcia:2012sz}. Inclusion of the likelihood constraining $\delta$ has no effect on the results, since it is only constrained by the background data. Note that a study applying model selection to neutrino oscillation data has been performed in \refcite{Bergstrom:2012yi}.

Electron spectra from beta decays of certain isotopes are sensitive to the kinematic effective mass-square
\be
m_\beta^2 \equiv \sum_i |U_{ei}|^2m_i^2 = m_1^2 c_{12}^2 c_{13}^2 + m_2^2 s_{12}^2 c_{13}^2 + m_3^2 s_{13}^2.
\ee
The results most  sensitive to $m_\beta^2$ are those of Mainz \cite{Kraus:2004zw} and Troitsk \cite{Aseev:2011dq}, yielding approximately Gaussian likelihoods
\bea
m_\beta^2 &=& -1.2 \pm 3.04 ~{\rm eV}^2, \\
m_\beta^2  &=& -0.67 \pm 2.53 ~{\rm eV}^2,
\eea
respectively.
\section{Parameter space and priors}
In this section we describe the space of parameters used and the priors imposed on them.
Since the likelihood of the tested data only depends on the particle physics parameters through $m_{ee}$, one could in principle 
perform the analysis using only that parameter, together with the NNEs and the nuisance parameters of the likelihood of \secref{sec:KKKanalysis}. However, since we want to take into account the non-trivial constraints from neutrino oscillation and tritium beta decay on $m_{ee}$, we instead choose to work with the full set of parameters of the Majorana mass matrix in terms of the masses and mixing parameters
\be\mbTh_{\rm{PF}} = (m_0, \dms, \dml, \theta_{12}, \theta_{23}, \theta_{13},\delta, \alpha, \beta ), \ee
where $m_0$ is the smallest neutrino mass, $\dms = m_2^2-m_1^2$, and $\dml = m_3^2-m_1^2$.  
Hence, the full set of $14$ parameters becomes 
\be \mbTh = (\mbTh_{\rm{PF}}, \NME_\Ge, \NME_\Xe, \rho_\Ge). \ee
This approach implies that the results in principle depend on the assumed mass ordering of the neutrinos, which can be either normal ($\dml >0$) or inverted ($\dml < 0$), but we will show that in practise the evidence of incompatibility between the data sets as well as the evidence for the existence of neutrinoless double beta decay does not, as long as the same priors are used for both mass orderings.

The purpose and result of including the background data is essentially to restrict the distribution of $m_{ee}$ which is then used to analyze the neutrinoless double beta decay data. Since the parameters $\dms, \dml, \theta_{12}, \theta_{23}$, and $\theta_{13}$ are very well-determined by oscillation data, their priors are rather irrelevant, and so we simply take uniform priors. 
Usually, one should not use the data to select priors for the parameters. In this case, however, one can get away with it as long as the posterior using $\Dbkg$ is the way we expect. The priors on the phases $\delta, \alpha, \beta$ are taken uniform, since this is the only choice consistent with the symmetries of the mass matrix \cite{Haba:2000be,Kass:1996}. 

The remaining particle physics parameter is $m_0$, which together with the two NMEs are those of interest. The priors on these parameters are also those for which the final inference may depend significantly.
We use two different priors on the lightest neutrino mass $m_0$. First, $m_0>0$ can be thought of as parameterizing the \emph{scale} of neutrino masses (at least down to $m_0 \simeq 10^{-2} \eV$ for the measured mass squared differences), and one could argue that it is most \qu{natural} for all the neutrino masses to be roughly of the same order of magnitude, or at least not differ by more than one or two orders. In this case, $m_0^2$ should not differ by many orders of magnitude from the mass squared differences, say $m_0 \simeq 10^{-3} - 1 \eV$. Hence, we take a log prior $m_0 \sim \LOG(10^{-3} \eV, \cdot)$,\footnote{The hard lower bound can again be avoided by using a prior which decays smoothly below $10^{-3}$ without altering the results.} which we call prior $\mathcal{A}$. Second, in order to consider the possibility of having more prior probability put on larger masses, we take a prior proportional to $1/\sqrt{m_0}$ with lower limit 0, which is denoted by $\mathcal{B}$. We do \emph{not} use a uniform prior on $m_0$, simply because we believe it to not reflect the view of the community.\footnote{One would have to believe that it is a priori equally probable for $m_0$ to be, say, in the interval $[0, 0.01]\eV$ as in the interval $[1,1.01]\eV$.}
Note that increasing the prior upper limit on $m_0$ does not change any results, since this region is anyway excluded by the background data. Equivalently, the corresponding reduction of the evidences will cancel in the Bayes factors.

\subsection{The nuclear matrix element uncertainties}\label{sec:NMEs}
If the nuclear matrix elements needed to predict the expected rate of neutrinoless double beta decay in \eref{eq:decayrate} were known accurately, combining and comparing the different data sets would definitely be easier. Of course, one could always perform such an analysis for fixed ratios of the NMEs. However, then it is difficult to specify how much those would be allowed to vary.
Instead, it is better if one can incorporate the NME uncertainty into the analysis from the very beginning.
But since the NMEs are in no way measured, one cannot include a factor in the likelihood constraining them, and, in general, an analysis of experiments using $J$ different nuclei have $J+1$ free parameters.

Incorporating the NME uncertainties in a Bayesian analysis is in principle straightforward, since one can simply use the NME calculations (as well as the properties of the different calculational methods) as prior information. This should yield at least a somewhat well-specified prior probability distribution of the NMEs. A unimodal prior with a width (which can be varied) parameterizing the uncertainty seems a natural choice. The NMEs are then included as free parameters and marginalized over in the end.

How does one then quantify the uncertainty in the NMEs? Models used to calculate the NMEs have constantly been improving leading to more accurate results, and compilations of recent representative calculations using different methods have been given, for example, in Refs.~\cite{GomezCadenas:2010gs,Rodejohann:2011mu}. In addition, there might be reasons, based on the properties of the methods and the assumptions used, to believe that some methods are likely to underestimate the NMEs, while some methods are probably overestimating them. 
This motivated the authors of \refcite{GomezCadenas:2010gs} do define \qu{physics-motivated} ranges of the NMEs for different nuclei. 


Furthermore, if many methods tend to under- or overestimate the NMEs in one nucleus, it is likely that they also do so in other. In other words, the NMEs of different nuclei should be positively correlated a priori. 
Most important for the comparison of experiments in two different nuclei is the ratio of NMEs, since a rescaling of both NMEs could be compensated by a change in $m_{ee}$ in \eref{eq:decayrate}. Since any biases in the calculations of the NMEs in different nuclei is expected to cancel in their ratio to some degree, it seems most straightforward to use the uncertainty in the ratio to specify the priors.

From Refs.~\cite{GomezCadenas:2010gs,Rodejohann:2011mu} we conclude that it is reasonable to assign $\NME_\Ge$ and $\NME_\Xe$ the same relative uncertainties, and we take the marginalized priors of the NMEs as log-normal, $\NME_\Xe \sim \LOGN(m_\Xe,\sigma_\NME)$ and $\NME_\Ge \sim \LOGN(m_\Ge,\sigma_\NME)$, with $m_\Xe = 2.8$ and $m_\Ge = 4.1$ as best estimates. If $\log \NME_\Xe$ and $\log \NME_\Ge$ then are jointly normally distributed with correlation coefficient $\rho$, then the ratio $r=\NME_\Xe/\NME_\Ge \sim \LOGN(m_\Xe/m_\Ge,\sigma_r)$ also has a log-normal prior with  $\sigma_r = \sigma_\NME^{\sqrt{2}\sqrt{1-\rho}}$. Hence, given as input the two uncertainties $\sigma_\NME$ and $\sigma_r$, one can calculate $\rho$ and the two-dimensional prior of the NMEs is defined.
The correlations between the different NMEs within one method of NME calculations have been studied in \refcite{Faessler:2008xj}, with which our correlations will roughly agree.

Following the discussion in \refcite{GomezCadenas:2010gs}, one could be optimistic regarding our current knowledge and take $\sigma_\NME =1.15$. The $95 \%$ central credible intervals for the NMEs are then $[3.1, 5.4]$ for $\NME_\Ge$ ($A=76$) and $[2.1,3.7]$ for $\NME_\Xe$ ($A=136$), which can be compared with Fig.~1 of \refcite{GomezCadenas:2010gs}. However, considering the possibility that such a small error is too optimistic, one should also take a more conservative value of the uncertainty such as $\sigma_\NME =1.3$, giving $95 \%$ central credible intervals as $[2.5, 6.9]$ and $[1.7, 4.7]$, respectively. The case of no NME uncertainty, $\sigma_\NME = 1$, is obviously unrealistic, but is included for comparison.

The ratio of the five estimates complied in \refcite{GomezCadenas:2010gs} lie in the range $r = 0.54 - 0.90$, while many of the ratios of \refcite{Simkovic:2009pp} are smaller, around $0.45$. In principle, assuming that the calculations of $\log r$ are independent of each other when conditioned on $r$\footnote{This is probably not true since some of the methods share common features.}, have the same errors, and are not biased in any direction, one could use Bayes theorem to obtain the posterior of $r$ (posterior to the calculations, but prior to the data).  With a log-normal density (uncertainty $\sigma_{\rm comp}$) of the calculations and with and a log prior on $r$, one obtains a log-normal posterior with median equal to the mean of the five calculations and uncertainty parameter $\sigma_r = \sigma_{\rm comp}^{1/\sqrt{5}}$. For $\sigma_{\rm comp} \simeq 1.25$ (consistent with the values of \refcite{GomezCadenas:2010gs}), $\sigma_r$ is only about $1.1$. However, since we realize that the assumptions going into this reasoning might not be completely valid, we also take a more conservative value $\sigma_r = 1.25$. The prior $95 \%$ central credible interval for $r$ is then $[0.44,1.06]$, we we think is wide enough. However, for comparison, we also leave room for even more conservatism, and sometimes also consider the uncertainties $(\sigma_\NME,\sigma_r) = (1.5, 1.35)$, for which the $95 \%$ central credible intervals for $\NME_\Ge$, $\NME_\Xe$, and $r$ are $[1.8, 9.1]$, $[1.3, 6.2]$, and $[0.38, 1.23]$, respectively.

\section{Results}\label{sec:results}
In this section we perform the combined analysis of the relevant data $D_\Ge, D_\Xe$, and $\Dbkg$ within the standard model with massive Majorana neutrinos. We especially emphasize the constraints and evidence from the neutrinoless double beta decay data $D_\Ge$ and $D_\Xe$.

However, before simply combining all the data to yield the final model and parameter inference, one should be convinced that the data are actually mutually consistent. We thus first want to test if the two sets of data in $\Dtest = (D_\Ge, D_\Xe)$ are consistent, when $\Dbkg$ is included as prior constraints. As discussed in \secref{sec:MScomptest}, the Bayes factor of compatibility vs.~incompatibility is 
\be \label{eq:bbcomtest}
\mathcal{R} =\frac{\Pr(D_\Ge, D_\Xe | \Dbkg, H)}{ \Pr(D_\Ge | \Dbkg, H)\Pr(D_\Xe | \Dbkg, H)} = \frac{\Pr(D_\Ge, D_\Xe, \Dbkg | H) \Pr(\Dbkg | H )}{\Pr(D_\Ge, \Dbkg | H)\Pr(D_\Xe, \Dbkg | H)}.
\ee
Note that any experiment yielding no evidence of neutrinoless double beta decay can never be \qu{more inconsistent} with $D_\Ge$ than the degree of which $D_\Ge$ is incompatible with the null hypothesis of no signal, regardless of which measure is used for the inconsistency. For the test of \eref{eq:bbcomtest} one would have
\be \mathcal{R} \geq \frac{\Pr(\DGe | H_0)}{\Pr(\DGe | \Dbkg, H)}. \label{eq:R_equality}\ee
When the upper limit on $m_{ee}$ from the data used to test $\DGe$ decreases, this inequality approaches an equality, as was discussed in \secref{sec:MScomptest}. As will be discussed in \secref{sec:decayev}, this lower limit is about $-5$ and $-6$ for $\log \mathcal{R}$ for priors $\mcA$ and $\mcB$, respectively.

In principle, one could also evaluate a Bayesian version of a p-value, encoding how \qu{extreme} or \qu{surprising} the observed data $D_\Ge$ is in the light of $D_\Xe$ (or the other way around). This can be done including the uncertainties of the model parameters (including parameters with only theoretical uncertainties).
For example, the probability that one would observe an equal or larger number of signal events $n_\Ge$ than was actually observed, given $D_\Xe$ and $\Dbkg$, is
\be \Pr(n_\Ge \geq n_\Ge^{\text{obs}} | D_\Xe,\Dbkg, H) = \int \Pr(n_\Ge \geq n_\Ge^{\text{obs}} | \mbTh, H) \Pr(\mbTh | D_\Xe, \Dbkg, H) \df^N \mbTh. \ee
Since the signal rate for a non-zero $m_{ee}$ is always larger than zero, it holds that $\Pr(n_\Ge \geq n_\Ge^{\text{obs}} | D_\Xe,\Dbkg, H) \geq \Pr(n_\Ge \geq n_\Ge^{\text{obs}} | H_0) \simeq 10^{-4}$, as was estimated in \secref{sec:KKKanalysis}. Hence, $\DGe$ can never be more surprising under $H$ than under $H_0$, no matter how small the upper limit on $m_{ee}$ is. As expected, we find larger p-values for the prior $\mcB$ on $m_0$ and for larger NME uncertainties. For all priors on $m_0$ and the NMEs (except the most conservative with $(\sigma_\NME,\sigma_r) = (1.5, 1.35)$), we obtain (for fixed background rate $b$) $\Pr(n_\Ge \geq n_\Ge^{\text{obs}} | D_\Xe, \Dbkg, H)$ between $2.8\cdot 10^{-4}$ ($3.4 \sigma$) and $1.8\cdot 10^{-3}$ ($2.9 \sigma$).
However, p-values are not directly related to the actual probability of the hypothesis being tested, and are not even the probability of the observed data, but a probability of data which has never been observed. 
Hence, rather than going to much into detail about these, we will concentrate on the consistency of the parameter constraints using the test based on model selection of \secref{sec:MScomptest}. 

If one fixes the ratio $r$ of the NMEs, both $\DGe$ and $\DXe$ essentially constrain only $m_{ee}$. In this case, one could perform rough frequentist hypothesis tests using the likelihood ratio. To test the compatibility, one can calculate the ratio of the full likelihood when maximized over $m_{ee}$ (and $\rho_{\Ge}$) to that obtained when maximizing the individual likelihoods separately.\footnote{This is essentially the method of \refcite{Maltoni:2003cu} extended to general likelihoods. The maximum of the Xe likelihood is taken at $m_{ee}=0$.} In fact, the ratio bears some resemblance to \eref{eq:bbcomtest}, but with the individual likelihoods maximized, rather than integrated, over the parameters (and the background data ignored). Although in this case we cannot know the precise distribution of the likelihood ratio, minus two times of its logarithm should asymptotically have a $\chi^2$-distribution with one degree of freedom. Converting the observed value of the likelihood ratio into a rough estimate of the significance yields the results of the first row of \tabref{tab:freqsig} for various values of $r$. As expected, the significance increases significantly when $r$ increases. Testing the hypothesis $m_{ee}$ by calculating the likelihood ratio similar to \eref{eq:Qsquare_Ge}, but now also including $\DXe$, yields the results in the second row of \tabref{tab:freqsig}, where the two values correspond to an assumed $\chi^2$-distribution with one and three degrees of freedom, respectively.\footnote{Of course, the same caveats as in \secref{sec:KKKanalysis} applies.}  As one observes, the results are very dependent on the chosen value of $r$. Hence, we prefer to concentrate on the Bayesian analysis, where the NME uncertainties are instead integrated over. 
\begin{table}
\begin{center}
\begin{tabular}{|c|c|c|c|c|c|c|c|}
\hline
$r = $ & $0.4$ & $0.5$ & $0.6$ & $0.7$  & $0.8$ & $0.9$  & $1.0$  \\ 
\hline
Against compatibility ($\sigma$'s) & $2.2$ & $2.6$ & $3.1$ & $3.4$ & $3.7$ & $3.9$ & $4.1$ \\ 
Against $m_{ee} = 0$ ($\sigma$'s) & $3.1/3.9$ & $2.8/3.6$ & $2.4/3.2$ & $2.0/2.9$  & $1.6/2.5$  & $1.2/2.1$ & $0.8/1.7$\\ 
\hline
\end{tabular}
\end{center}
\caption{Rough estimates of the significance when testing the compatibility of the two neutrinoless double beta decay data sets for different fixed values of the NME ratio $r$, and the significance when testing $m_{ee} = 0$ against $m_{ee}>0$ for the combination of data.\label{tab:freqsig}}
\end{table}

\begin{figure}[t]
\centering
\includegraphics[width=0.7\textwidth]{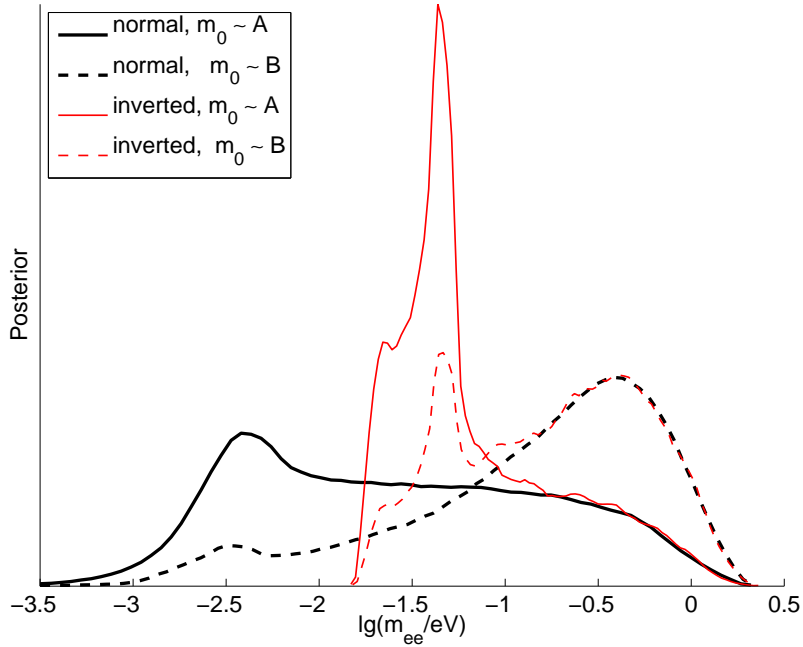}
\caption{Posteriors using only background data of the logarithms of $m_{ee}$ for different priors and assumed mass orderings.}\label{fig:bkg_posterior}
\end{figure}

The different evidences in \eref{eq:bbcomtest} can be associated with different posterior distributions: the posteriors using i) only background data, ii) background together with one set of tested data,  iii) background together with the other set of tested data, and iv) background together with both sets of tested data. The posteriors using only the background data of the logarithm of $m_{ee}$ for both priors on $m_0$ and both mass orderings are displayed in \figref{fig:bkg_posterior}. All the posteriors are correctly normalized with respect to each other. Note that the posterior of $m_{ee}$ is independent of the mass ordering for $m_{ee} \gtrsim 0.1 \eV$. The peaks appearing for small $m_{ee}$ are due to the \qu{band structure} of the constraints in the $m_0$ - $m_{ee}$ plane (see, \eg, Fig.~2 of \refcite{Bilenky:2012qi}). Putting more prior probability on small values of $m_0$ will result in these peaks 
being more pronounced, but that $m_{ee}$ is always bounded from below (in a probabilistic sense) for both mass orderings.

\begin{figure}[t]
\centering
\includegraphics[width=0.49\textwidth]{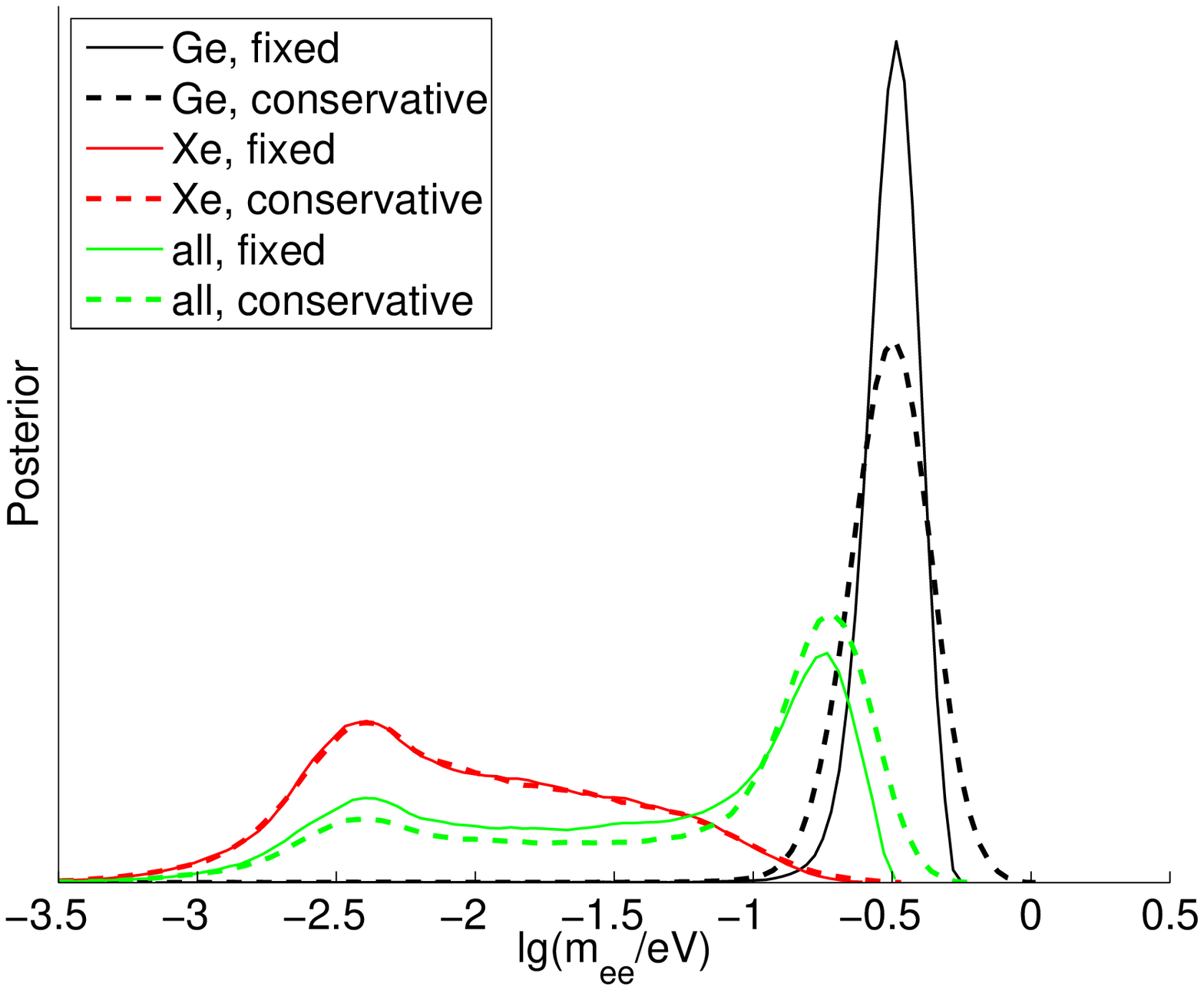}
\includegraphics[width=0.49\textwidth]{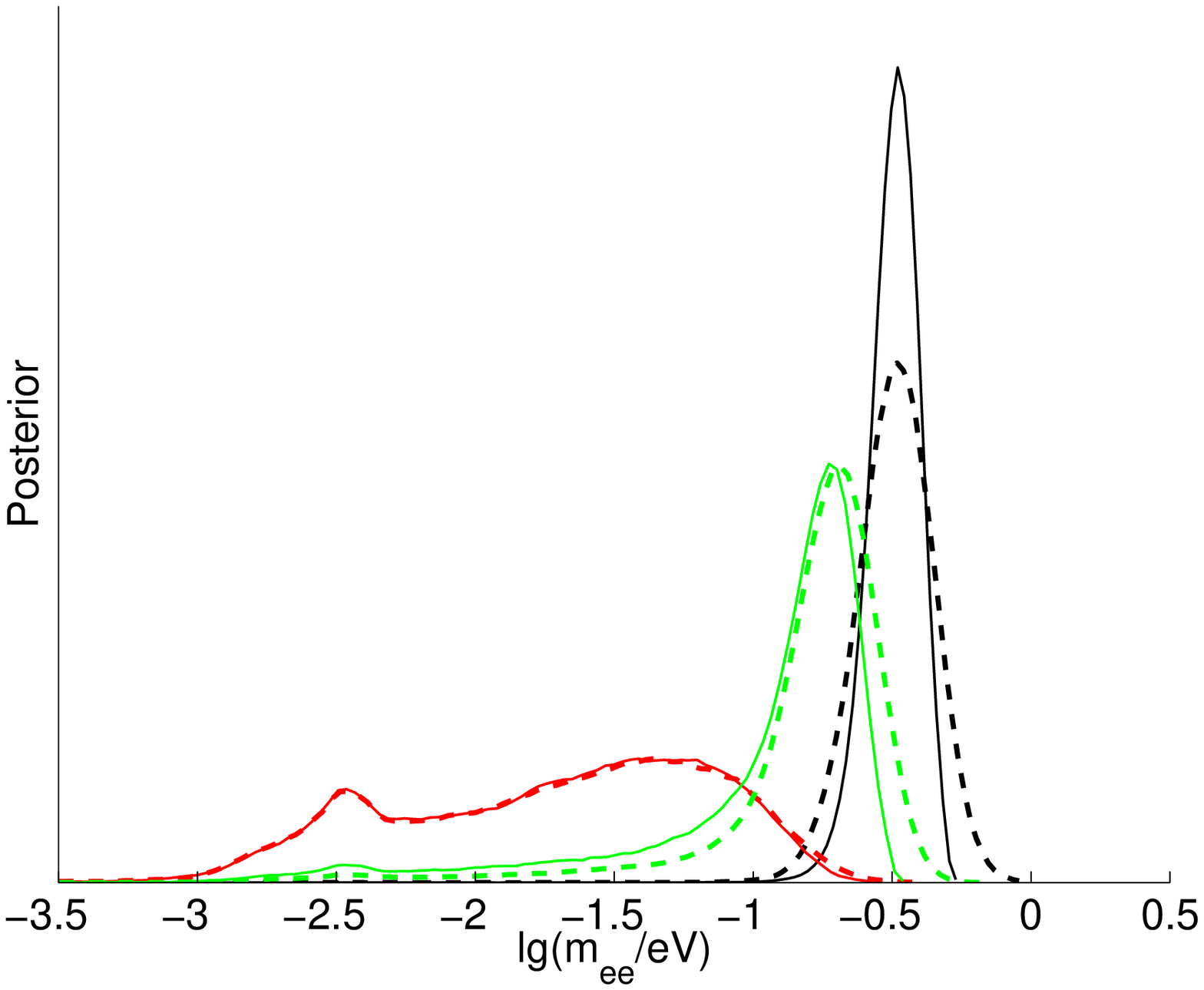}
\caption{Posteriors for the logarithm of $m_{ee}$ for normal mass ordering for different data sets and for both fixed NMEs and conservative priors with $(\sigma_\NME, \sigma_r) = (1.3, 1.25)$. The prior on $m_0$ is $\mcA$ (left) and $\mcB$ (right) and $\Dbkg$ is always used.}\label{fig:all_posterior}
\end{figure}

In the same manner, the posteriors for normal mass ordering\footnote{For the inverted mass ordering, the posteriors differ in the region of small $m_{ee}$, but are essentially identical in the interesting region of large $m_{ee}$.} when also using $\DGe$, $\DXe$, and all data, are shown in \figref{fig:all_posterior}, with $m_0 \sim \mcA$ in the left panel and $m_0 \sim \mcB$ in the right, and for the case of fixed NMEs and the prior with $(\sigma_\NME, \sigma_r) = (1.3, 1.25)$. $95 \%$ central credible intervals for $m_{ee}$ using $\DGe$ is $[0.22 \eV, 0.42 \eV]$ for fixed NMEs and $[0.18 \eV, 0.51 \eV]$ for $\sigma_\NME = 1.3$, which on a logarithmic scale is about a factor of $1.6$ larger. Hence, the statistical and NME uncertainties are roughly of the same size. One notes that there is some overlap between the posteriors using $\DXe$ and $\DGe$. However, it is only the marginalization down to one dimension which is visible, and this does not tell you how much the posteriors in the full parameter space are overlapping. There are posterior correlations between all the three parameters $m_{ee}$ (or $m_0$), $\NME_\Ge$ and $\NME_\Ge$, since there is an a posteriori correlation between $m_{ee}$ and the NME corresponding to the data used, and since that NME is also a priori correlated with the other NME. The test using \eref{eq:bbcomtest} takes into account all the constraints in the full parameter space.

\begin{figure}[t]
\centering
\includegraphics[width=0.7\textwidth]{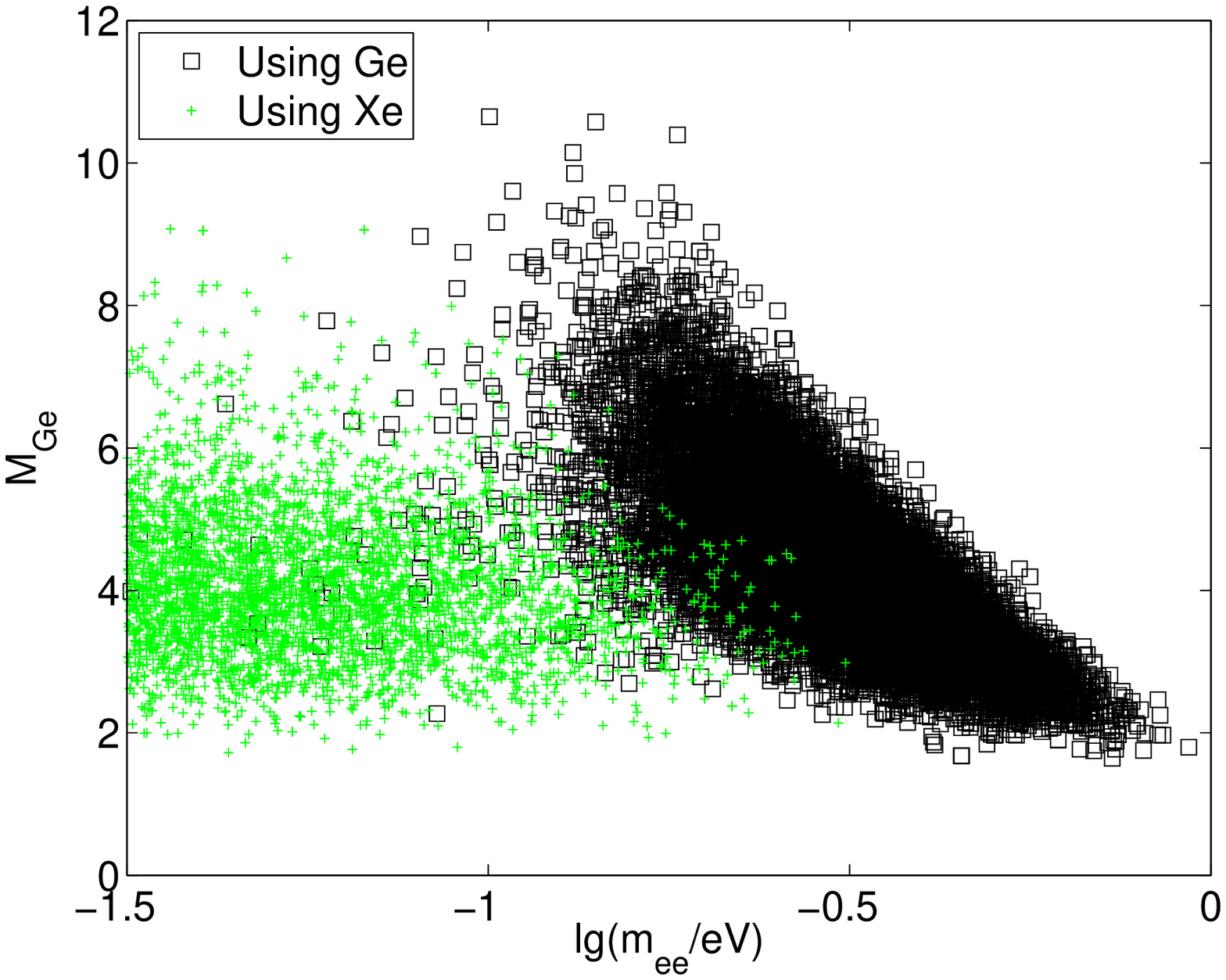}
\includegraphics[width=0.7\textwidth]{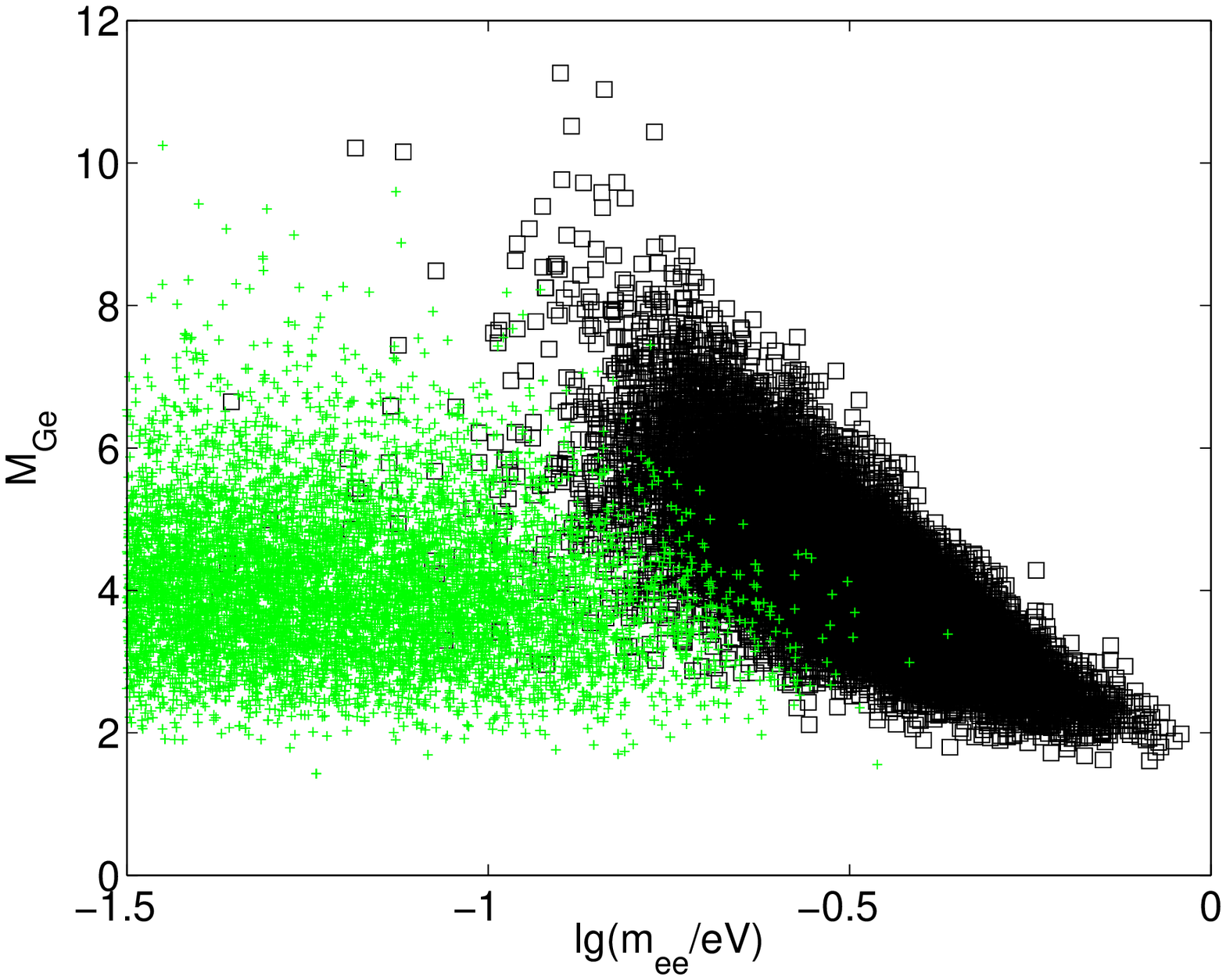}
\caption{Equally weighted samples from the posterior using $\DGe$ (green pluses) and $\DXe$ (black squares) for the prior $\mcA$ (top) and $\mcB$ (bottom) on $m_0$ and  $ (\sigma_\NME, \sigma_r) = (1.3,1.25)$. In total 20000 points were drawn, but some points are outside the plot ranges.}\label{fig:2Dpost}
\end{figure}

To show the constraints in two dimensions, we draw $20 000$ points from the full posterior and show their distribution in the $(m_{ee}, \NME_\Ge)$ plane in \figref{fig:2Dpost}, using $m_0 \sim \mcA$ (top) and $m_0 \sim \mcB$ (bottom) and $(\sigma_\NME, \sigma_r) = (1.3, 1.25)$.  In order to increase readability, points below $m_{ee} \simeq 0.03 \eV$ are not shown. One can see that there is a small amount of posterior using $\DXe$ (green pluses) that is overlapping with the region preferred by $\DGe$ (black boxes) around $m_{ee} \simeq 0.3 \eV$, as well as some posterior using $\DGe$ which is found in the region $m_{ee} \lesssim 0.1 \eV$.  The posteriors in the $(m_{ee}, \NME_\Xe)$ plane are quite similar.
There is a slightly larger overlap in the high-$m_{ee}$ region for $m_0 \sim \mcB$, while there is a smaller overlap in the region $m_{ee} \lesssim 0.1 \eV$. These effects will party cancel each other, making the compatibility test less dependent on the prior on $m_0$.

\begin{table}
\begin{center}
\begin{tabular}{|c|c|c|c|c|c|}
\hline
$(\sigma_\NME, \sigma_r) = $ & $(1,1)$ & $(1.15,1.1)$ & $(1.3,1.1)$ & $(1.3,1.25)$  & $(1.5,1.35)$  \\ 
\hline
$m_0 \sim \mcA$ & $-4.42$ & $-4.31$ & $-4.31$ & $-4.06$ & $-3.66$  \\ 
$m_0 \sim \mcB$ & $-4.42$ & $-4.35$ & $-4.28$ & $-3.68$  & $-3.28$  \\ 
\hline
\end{tabular}
\end{center}
\caption{Calculated values of the logarithms of the Bayes factor $\mcR$ for different priors on $m_0$ and the NMEs. All statistical errors on the numerical estimates are $0.07$.}
\label{tab:Rvalues}
\end{table}

Our results for the logarithm of the Bayes factor $\mathcal{R}$ are summarized in \tabref{tab:Rvalues}, which all have statistical errors of $0.07$ at one standard deviation. In general, the obtained values only depends weakly on the choice of priors within the set we consider, and is between moderate and strong against compatibility, according to Jeffrey's scale. As expected, the evidence is independent of the absolute NME uncertainties, but does depend on the uncertainty of the ratios of the NMEs.
Furthermore, using prior $\mcB$ yields smaller evidence against compatibility, but only slightly so, and only for the larger NME ratio uncertainty. Fixing the NMEs is clearly to restrictive, but essentially equivalent results are obtained for $ \sigma_r = 1.1$. In addition, the evidence against compatibility does not decrease too much when using the very large NME uncertainties $ (\sigma_\NME, \sigma_r) = (1.5,1.35)$.
In total, considering the most realistic cases by excluding both the results for fixed NMEs and for the largest uncertainties, we obtain
\be \quad  \log \mathcal{R} \in [-4.35,-3.68],   \quad  \mathcal{R}^{-1} \in [40, 80].   \ee 
Hence, one can say that the data $\DGe$ and $\DXe$ are about 40 to 80 times more probable under the hypothesis that they are incompatible, and hence need different sets of parameters to describe the data. If one assigns equal prior probabilities to the two hypotheses, one obtains
\be \Pr(C|\DGe,\DXe,\Dbkg) \simeq 1.3 \% - 2.5 \%.\ee
Lowering the limit on $m_0$ for the log prior $\mcA$ will decrease the evidence against compatibility, but for reasonable lower limits (say, above $10^{-6}$), this effects is very small. In addition, one can consider putting some small but finite prior probability at $m_0 = 0$, but since this will also only change the results marginally we do not consider this possibility further.

In the end of \secref{sec:KKKanalysis}, it was noted that the efficiency $\epsilon$ might be smaller than one, but that we have simply included $\epsilon =1 $ as part of the claim of \refcite{KKK}. A smaller $\epsilon$ would require larger $m_{ee}$ to fit the data, which would essentially mean that the black curves and points of Figs.~\ref{fig:all_posterior} and \ref{fig:2Dpost} would be shifted $\lg(\epsilon)/2$ log-units to the right, and decreasing $\epsilon$ can thus only increase the degree of incompatibility with $\DXe$, independent of how that incompatibility is evaluated. As a result, the evidence of the existence of decay for the full set of data is expected to decrease.

We have tested the compatibility of the different sets of neutrinoless double beta decay experiments with $\Dbkg$ used as prior constraints on the model parameters. In principle, one could also try to perform the same test while ignoring the background data. In this case, however, the prior on $m_{ee}$ would be very difficult to specify, and this would make the results of the compatibility tests very prior dependent. However, we think that the relevant question is instead what we have investigated in this work, \ie, the consistency of the different neutrinoless double beta decay data sets, within the standard model with massive Majorana neutrinos together with all its already existing constraints.

Finally, we note that we have checked that we obtain reasonable results if various inputs to our analysis are changed. For example, removing the constraints of $\DXe$ by fixing $\NME_\Xe$ to a very small value, we find that the evidence against compatibility disappears ($\mathcal{R} \simeq 1$), and if we remove the signal from \figref{fig:spec}, we obtain a value of $\mathcal{R}$ just above unity. If we instead inject a signal in $\DXe$ consistent with $m_{ee} \simeq  0.3 \eV$ and $\NME_\Xe \simeq 2.8$ (with similar width as the real likelihood), we do obtain weak to moderate evidence in favor of compatibility.

\subsection{Evidence of the decay from different data sets}\label{sec:decayev}
One can also evaluate the evidence of the decay from $\DGe$ as in \secref{sec:KKKanalysis} but within the full model with massive neutrinos, \ie, with the signal strength being derived from the particle physics parameters and  $\NME_\Ge$. Again, using $\Dbkg$ to constrain the model parameters, one obtains the Bayes factor
\be \mathcal{B}_{\text{Ge}} = \frac{\Pr(\DGe | H, \Dbkg )}{\Pr( \DGe  | H_0, \Dbkg)} =  \frac{\Pr(\DGe,\Dbkg | H )}{\Pr(\DGe  | H_0) \Pr(\Dbkg | H)}, \ee
since the conditioning on $\Dbkg$ is irrelevant under $H_0$. We find $\log \mathcal{B}_{\text{Ge}} \simeq 5.2 ~(6.0) $ for the prior $\mcA$ ($\mcB$).\footnote{All logarithms of Bayes factors in this section have a numerical uncertainty of 0.1 or smaller.}
Once again, the inclusion of the background data eliminates the dependence on the upper limit on the prior on $m_0$. These are very similar to the values found in \tabref{tab:logB_KKK} when the signal strength $s$ was the free parameter. It is, as expected, independent of the NME uncertainties and the assumed mass ordering. The latter is because, although the ranges of the priors on $m_{ee}$ are different, the priors in the region $m_{ee}>0.1 \eV$ are the same (see \figref{fig:all_posterior}) for both mass orderings.

Although there is substantial evidence against the data sets being compatible, one can disregard this fact and still calculate the total evidence of neutrinoless double beta decay.
When all experiments are considered, we obtain the Bayes factor
\be \mathcal{B}_{\text{tot}} = \frac{\Pr(\DXe,  \DGe | H, \Dbkg )}{\Pr(\DXe,  \DGe  | H_0, \Dbkg)} =  \frac{\Pr(\DXe,  \DGe,\Dbkg | H )}{\Pr(\DXe,  \DGe  | H_0) \Pr(\Dbkg | H)}. \ee
For all combinations of priors on $m_0$ and the NMEs (excluding the most conservative NME priors), we find $\log \mathcal{B}_{\text{tot}} = 0.5 - 0.8$, which is no evidence or only very weak. The exception is for $m_0 \sim \mcB$ and $(\sigma_\NME,\sigma_r) = (1.3, 1.25)$, for which $\log \mathcal{B}_{\text{tot}} \simeq 1.33$. The reason for this is that the larger NME errors allows $\DGe$ and $\DXe$ to be better fitted simultaneously while the prior on $m_0$ puts more prior probability in the region preferred by the combination of data. 
Our rough estimates of the frequentist significances for fixed NME ratios was discussed earlier and are summarized in \tabref{tab:freqsig}.

If one believes that the evidence against compatibility between the data sets is best treated by simply ignoring the claim of \refcite{KKK}, one obtains the evidence for neutrinoless double beta decay as
\be \mathcal{B}_{\Xe} = \frac{\Pr(\DXe | H, \Dbkg )}{\Pr(\DXe | H_0, \Dbkg)} =  \frac{\Pr(\DXe,\Dbkg | H )}{\Pr(\DXe,   | H_0) \Pr(\Dbkg | H)}. \ee
We find $ \log \mathcal{B}_{\text{Ge}} \simeq -0.3 ~ (-0.95) $ for the for prior $\mcA$ ($\mcB$). In other words, prior $\mcB$ puts more prior in the large signal regions, which, when these are subsequently excluded by the data, leads to stronger evidence against the signal hypothesis.
\section{\label{sec:conclusions}Summary and conclusions}
Due to the large theoretical uncertainties of the calculated nuclear matrix elements, a statistical analysis of neutrinoless double beta decay experiments within the standard model with massive Majorana neutrinos, or any other model predicting the decay, is not straightforward.
We have chosen to perform a Bayesian analysis, which, in a addition to all the usual advantages, makes it possible to take these uncertainties into account in a statistically coherent manner. 

From the analysis of the data used to claim the observation of neutrinoless double beta decay in \Gew~we find strong evidence in favor of a peak in the spectrum and moderate evidence that the peak is actually close to the energy expected for the neutrinoless decay. We also find a lower significance and a significantly larger statistical error than the original analysis, which we have taken into account when comparing with the other data.

Before one combines all the data to yield the final constraints on the models and their parameters, one should first test if the different data sets are mutually compatible. We have performed such a test of the consistency of the claim in \Gew~with the recent measurements using \Xew, within the standard model with massive Majorana neutrinos, and found that the two data sets are about 40 to 80 times more probable under the assumption that they are incompatible, depending on the nuclear matrix element uncertainties and the prior on the lightest neutrino mass. In other words, there is moderate to strong evidence of incompatibility, and for equal prior probabilities, the posterior probability of compatibility is between $1.3 \%$ and $2.5 \%$. The results are only weakly dependent on the choice of priors and NME uncertainties.
If one, despite such evidence for incompatibility, combines the two measurements, we find that there is no significant evidence of neutrinoless double beta decay. If one ignores the claim, there is weak evidence against the existence of the decay. 
We have also performed approximate frequentist tests of the compatibility of the two sets of experiments, assuming different fixed ratios of the nuclear matrix elements, and have found that the results depend strongly on the value of the NME ratio, as expected.

In addition to the experiments using \Xew~as the decaying nucleus used in this work, there are other experiments utilizing other nuclei expected to deliver data in the near future. This includes GERDA \cite{Smolnikov:2008fu} which is searching for the decay of \Gew, enabling a comparison with the claim of \refcite{KKK} without needing to consider the NME uncertainties. In order get the most information out of the experiments on possible models generating the decay, combined analyses within those models should be performed. The analysis of this work could then be extended to other combinations of experiments, as well as different mediating mechanisms. If GERDA were to find evidence of neutrinoless double beta decay, while Xenon-based experiments continue to decrease the upper limits on the \Xew~decay rate, one would have to consider other particle physics models as possible sources of the decay, and an extension of the Bayesian analysis presented here could be used to differentiate between such models.

\appendix


\section*{\label{sec:Ack}\noindent{Acknowledgments}}
The author would like to thank B.~Schwingenheuer and N.~V.~Karpenka for helpful comments.

\bibliographystyle{JHEP}
\bibliography{0nbb_cons}

\end{document}